%% file: SciDef.tex
\documentclass[sigconf,preprint,natbib=true]{acmart}

\acmDOI{}
\acmISBN{}
\usepackage[utf8]{inputenc} % remove if using xelatex/lualatex
\usepackage{microtype}
\usepackage{graphicx}
\usepackage{algorithm}
\usepackage{tabularx}
\usepackage{amsfonts}
\usepackage{subcaption} % loads caption; keep single caption stack to avoid extra vertical space
\usepackage{cleveref}
\usepackage[table]{xcolor} % provides \textcolor and \arrayrulecolor (via colortbl)
\usepackage{bbm}
\usepackage{enumitem}

\definecolor{taxobg}{HTML}{EFEFEF}
\definecolor{oursbg}{HTML}{EAF3FF}

\title[SciDef]{SciDef: Datasets and Tools for Automated Definition Extraction from Scientific Literature with LLMs}

\author{Filip Ku\v{c}era}
\affiliation{%
  \institution{National Institute of Informatics (NII)}
  \city{Tokyo}
  \country{Japan}}
\affiliation{%
  \institution{University of W\"urzburg}
  \city{W\"urzburg}
  \country{Germany}}
\email{f.kucera@media-bias-research.org}

\author{Christoph Mandl}
\affiliation{%
  \institution{National Institute of Informatics (NII)}
  \city{Tokyo}
  \country{Japan}}
\affiliation{%
  \institution{University of Passau}
  \city{Passau}
  \country{Germany}}
\email{c.mandl@media-bias-research.org}

\author{Isao Echizen}
\affiliation{%
  \institution{National Institute of Informatics (NII)}
  \city{Tokyo}
  \country{Japan}}
\email{iechizen@nii.ac.jp}

\author{Radu Timofte}
\affiliation{%
  \institution{University of W\"urzburg (JMU)}
  \city{W\"urzburg}
  \country{Germany}}
\email{radu.timofte@uni-wuerzburg.de}

\author{Timo Spinde}
\affiliation{%
  \institution{National Institute of Informatics (NII)}
  \city{Tokyo}
  \country{Japan}}
\email{t.spinde@media-bias-research.org}

\renewcommand{\shortauthors}{Ku\v{c}era et al.}

\copyrightyear{2026}
\acmYear{2026}
\acmConference[CIKM '26]{The 35th ACM International Conference on Information and Knowledge Management}{November 7--11, 2026}{Rome, Italy}

\begin{document}

%\begin{abstract}

%Definitions are the foundation for any scientific work, yet extracting definitions for a given keyword across the literature is increasingly difficult as publication volume grows. We therefore release two resources for scientific definition extraction: \textit{DefExtra}, a dataset of human-extracted definitions from academic papers, and \textit{DefSim}, a dataset of human-annotated definition-pair similarity judgments. To validate and support reuse of these datasets, we also release \textit{SciDef}, a reproducible LLM-based pipeline for definition extraction. Using SciDef, we evaluate 16 language models under multiple prompting strategies and show that multi-step and DSPy-optimized prompting improve extraction performance. The best configuration extracts $86.4\%$ of definitions from \textit{DefExtra}. We further compare evaluation metrics for LLM-extracted definitions and find that an NLI-based method yields the most reliable results and achieves high agreement with human judgments on \textit{DefSim}. We show that LLMs can recover many definitions from scientific literature, while also exposing a key limitation: models often over-generate definitions. Future work should therefore focus on relevance-aware extraction and filtering. Code \& datasets are available at \href{https://github.com/Media-Bias-Group/SciDef}{https://github.com/Media-Bias-Group/SciDef}.

%\end{abstract}

\begin{abstract}
Scientific concepts are often defined inconsistently across papers, making it difficult to compare findings, reuse terminology, and build reliable downstream resources. We present \textsc{SciDef}, a resource suite for scientific definition extraction. The suite contains \textsc{DefExtra}, a benchmark of 268 human-validated author-stated definitions from 75 academic papers; \textsc{DefSim}, 60 human-labeled definition-pair similarity judgments; and an open LLM-based pipeline for PDF preprocessing, chunking, definition extraction, prompt optimization, and evaluation. We validate the resources by benchmarking 16 language models across prompting strategies and chunking schemes. The strongest set-level configuration achieves a score of 0.397, while the highest-coverage configuration matches at least one prediction to 86.4\% of gold definitions but over-generates candidate definitions. We further show that an NLI-based matching metric agrees strongly with human DefSim judgments. These results position SciDef as a reusable benchmark and tooling layer for definition-centric literature analysis, while highlighting relevance-aware filtering as the key bottleneck for fully automatic definition extraction. Code \& datasets are available at \href{https://github.com/Media-Bias-Group/SciDef}{https://github.com/Media-Bias-Group/SciDef}.
\end{abstract}

\begin{CCSXML}
<ccs2012>
   <concept>
       <concept_id>10002951.10003227.10003351</concept_id>
       <concept_desc>Information systems~Data mining</concept_desc>
       <concept_significance>300</concept_significance>
       </concept>
   <concept>
       <concept_id>10002951.10003317.10003347.10003352</concept_id>
       <concept_desc>Information systems~Information extraction</concept_desc>
       <concept_significance>500</concept_significance>
       </concept>
   <concept>
       <concept_id>10010147.10010178.10010179</concept_id>
       <concept_desc>Computing methodologies~Natural language processing</concept_desc>
       <concept_significance>300</concept_significance>
       </concept>
   <concept>
       <concept_id>10010147.10010178.10010179.10010186</concept_id>
       <concept_desc>Computing methodologies~Language resources</concept_desc>
       <concept_significance>500</concept_significance>
       </concept>
 </ccs2012>
\end{CCSXML}

\ccsdesc[300]{Information systems~Data mining}
\ccsdesc[500]{Information systems~Information extraction}
\ccsdesc[300]{Computing methodologies~Natural language processing}
\ccsdesc[500]{Computing methodologies~Language resources}

\keywords{definition extraction, large language models, scientific literature mining, evaluation metrics, natural language inference, semantic similarity}

\maketitle

\input{latex/Sections/Introduction}

\input{latex/Sections/RelatedWork}

\input{latex/Sections/Methodology}

\input{latex/Sections/Results}

% \subfile{latex/Sections/Limitations}

\input{latex/Sections/Conclusion}

\input{latex/Sections/Acknowledgment}

\input{latex/Sections/GenAIDisclosure}

% \FloatBarrier

% --- Bibliography (ACM requires the style line) ---
\bibliographystyle{ACM-Reference-Format}
\bibliography{CustomRebiber2}

% \subfile{latex/Sections/Appendix}

\end{document}

%% file: latex/Sections/Introduction.tex
\section{Introduction}

Definitions are foundational to scientific work because they determine how concepts are introduced, interpreted, compared, and reused across studies \cite{taxocom_2022, sas_capiluppi_2024_taxonomy, taxoenrich_2022}. Without shared conceptual structures, research can become fragmented and difficult to reproduce \cite{Diaba-Nuhoho2021}. However, as publication volumes grow, manually finding and consolidating definitions for a given keyword becomes increasingly infeasible \cite{fodor-etal-2025-compositionality}. Motivated by this growing problem, we propose developing reusable resources for extracting, comparing, and evaluating scientific definitions from academic literature.

For definition extraction, large language models (LLMs) are promising tools, as they can recognize definitional statements beyond surface-level keyword patterns~\cite{Spinde2025}. Yet, reproducible research on the extraction of scientific definitions using LLMs remains limited. Existing work lacks a public benchmark of definitions extracted from academic papers, a dedicated dataset for evaluating definition similarity, and reusable infrastructure for comparing extraction pipelines and prompting strategies \cite{banerjee_2024_few-shot, sun2023discoveringpatternsdefinitionsmethods, Spinde2025TaxoMatic, xuSurveyTerminologyExtraction2025a, taxoenrich_2022, taxocom_2022}.

Therefore, we release a resource suite for extracting and evaluating scientific definitions. \textit{DefExtra} is a benchmark of $268$ human-extracted definitions from $75$ curated academic papers, including context spans, metadata, and explicit/implicit definition labels. \textit{DefSim} contains $60$ definition pairs with human semantic-similarity labels for validating definition-matching metrics. \textit{SciDef} is a reproducible LLM pipeline that supports extractor and metric comparison, as well as definition collection for literature reviews, taxonomy construction, ontology engineering, and domain mapping.

We evaluate the resource suite using media bias as a challenging case study, since the domain is widely studied but inconsistently defined across disciplines~\cite{spinde_media_2023}. Using \textit{SciDef}, we benchmark $16$ LLMs under multiple prompting strategies, including one-step, multi-step, few-shot, and DSPy-optimized prompting \cite{khattab2023dspycompilingdeclarativelanguage}. We also compare similarity metrics and show that an NLI-based method aligns reliably with human judgments from \textit{DefSim}. Our experiments show that LLMs can recover many scientific definitions, with the highest-recall configuration extracting $86.4\%$ of definitions from \textit{DefExtra}. At the same time, models often over-generate candidate definitions, highlighting relevance-aware filtering as a key open challenge.

The key contributions of this resource paper are:
\begin{itemize}
    \item \textbf{DefExtra}, a benchmark of 268 human-extracted definitions from 75 academic papers, including source metadata, definition type labels, out-of-domain indicators, and copyright-protecting metadata for context reconstruction.
    \item \textbf{DefSim}, a task-specific validation resource of 60 definition-pair similarity judgments for evaluating semantic matching metrics on extracted scientific definitions.
    \item \textbf{SciDef}, an open, documented, and versioned LLM-based pipeline for PDF preprocessing, chunking, definition extraction, prompt optimization, and evaluation.
    \item \textbf{Resource validation experiments} showing how the released datasets and pipeline support comparison across 16 LLMs, multiple prompting strategies, chunking schemes, and semantic matching metrics.
\end{itemize}

%% file: latex/Sections/RelatedWork.tex
\section{Related Work} \label{sec:related}

Definitions enable consistent communication and reuse of scientific concepts \cite{Diaba-Nuhoho2021}, but terminological consensus is difficult in many fields \cite{Liesefeld2024, Kagan2024}. The domain of media bias illustrates this problem: prior work defines biases such as partisan reporting \cite{Groseclose2005}, linguistic bias \cite{spinde-etal-2021-neural-media}, or bias by word choice \cite{hamborg2019automated}, which describe similar concepts under different names, limiting comparability \cite{spinde_media_2023}. Systematic reviews and content analyses can consolidate such definitions, but they are labor-intensive and difficult to scale \cite{krippendorff2019}.

To find solutions to this problem, prior definition extraction work includes structured systems such as DefIE \cite{bovi2015defie}, sequence-labeling models \cite{Veyseh_Dernoncourt_Dou_Nguyen_2020}, term-extraction and prompt-engineering approaches \cite{banerjee_2024_few-shot,vatsal2024surveypromptengineeringmethods,xuSurveyTerminologyExtraction2025a}, and broader ontology-learning methods \cite{maedche2001ontology}. Existing resources include WCL for Wikipedia-based definition and hypernym extraction \cite{navigli2010annotated,navigli-velardi-2010-learning}, W00 for term/definition labels in ACL workshop papers \cite{jin-etal-2013-mining}, DEFT for free and semi-structured text from textbooks and SEC filings \cite{spala-etal-2019-deft}, and dictionaries such as Oxford and Urban Dictionary \cite{oxford_dictionary,urban_dictionary}. TaxoMatic uses LLM prompting for taxonomy-oriented definition extraction \cite{Spinde2025TaxoMatic}. 
% We build on this line of work by releasing a validated academic paper benchmark, task-specific definition-similarity labels, and a reusable LLM-based extraction pipeline, with TaxoMatic serving as the baseline.

Finally, to compare contents such as definitions, existing work explored various semantic similarity metrics. Transformer embeddings and LLM representations improve over lexical matching \cite{agirre2012semeval,reimers2019sbert,gao2021simcse,brown2020language}, while NLI models capture entailment-style relations useful for asymmetric or partial matches \cite{bowman2015snli,williams2018multinli,chen-etal-2022-semeval}. Benchmarks such as STS3k \cite{fodor-etal-2025-compositionality}, SICK \cite{marelli2014sick}, MSRP \cite{doland2005paraphrase}, and QQP \cite{sharma2019nlu} support general metric selection, but do not evaluate scientific definition pairs. \textit{DefSim} provides such task-specific validation data.

%% file: latex/Sections/Methodology.tex
\section{Resource Construction and Evaluation} 
\label{sec:methodology}

Our resources comprise two datasets, \textit{DefExtra} and \textit{DefSim}, and the LLM-based \textit{SciDef} pipeline. \textit{DefExtra} benchmarks scientific-definition extractors; \textit{DefSim} validates definition-matching metrics; and \textit{SciDef} provides a reproducible workflow for PDF preprocessing, chunking, LLM extraction, prompt optimization, and evaluation. % Together, they support extractor/metric comparison and candidate-definition collection for literature reviews, taxonomy construction, ontology engineering, and domain mapping.

\paragraph{Availability, licensing, ethics, and provenance.}
Following FAIR~\cite{Wilkinson2016FAIR}, \textit{DefExtra}, \textit{DefSim}, and \textit{SciDef} are archived with persistent DOIs~\cite{DefExtraZenodo,DefSimZenodo,SciDefZenodo} and released through Zenodo, Hugging Face, and GitHub; \textit{SciDef} is released under Apache-2.0, and \textit{DefExtra}/\textit{DefSim} under CC BY 4.0. To reduce copyright risk in \textit{DefExtra}, we release paper identifiers, metadata, and hydration scripts instead of source PDFs. The datasets contain no private or sensitive information. Human annotation was conducted by authors and academic collaborators; under the authors' institutional guidelines, this work did not require IRB review. Provenance, schemas, hydration instructions, full appendix, and run configurations are documented in the released repositories.

\begin{table}
\centering
\scriptsize
\setlength{\tabcolsep}{2.5pt}
\renewcommand{\arraystretch}{1.04}
\begin{tabularx}{\columnwidth}{@{}p{1.3cm}p{1.25cm}X@{}}
\toprule
\textbf{Source} & \textbf{Size} & \textbf{Input $\rightarrow$ Label(s)} \\
\midrule

\multicolumn{3}{@{}l}{\textbf{\textit{DefExtra}: extraction benchmark}} \\

\rowcolor{taxobg}
\cite{Spinde2025TaxoMatic} &
$113$ papers / $123$ defs. &
Paper ID $\rightarrow$ author-stated and annotator-synthesized definitions \\

\rowcolor{oursbg}
This work &
$75$ papers / $268$ defs. &
Paper ID $\rightarrow$ term, context, definition, exp./imp. type, metadata, OOD flag \\

\midrule

\multicolumn{3}{@{}l}{\textbf{\textit{DefSim}: definition similarity benchmark}} \\

\rowcolor{oursbg}
This work &
$60$ pairs &
Definition pair + context $\rightarrow$ human label ($1$--$5$), NLI score/label ($1$--$5$) \\

\bottomrule
\end{tabularx}
\caption{Overview of the two released datasets. The source column and color coding distinguish prior work from this work; OOD = out-of-domain.}
\Description{A table comparing the prior TaxoMatic resource with DefExtra and DefSim by source, size, input format, and labels.}
\label{tab:dataset}
\end{table}

\subsection{DefExtra: Definition Extraction Benchmark}
\label{sec:meth:defextra}

\textit{DefExtra} is a benchmark for extracting author-stated definitions from academic papers. It focuses primarily on media bias, a domain where definitions are diverse, contested, and spread across disciplinary boundaries~\cite{SPINDE2021102505}. As summarized in \Cref{tab:dataset}, \textit{DefExtra} builds on the TaxoMatic annotation effort~\cite{Spinde2025TaxoMatic}, which provided a media-bias-centered search space, Semantic Scholar retrieval, GROBID preprocessing, and first-pass annotations. In TaxoMatic, terms from an existing media-bias taxonomy~\cite{spinde_media_2023} were expanded with GPT-3.5-turbo, open-access PDFs were parsed with GROBID~\cite{GROBID}, and six trained annotators\footnote{Chosen among past or present collaborators with academic media-bias domain expertise.} screened highly cited papers for relevance and extracted initial definitions from relevant full texts.

This work builds on these initial annotations and builds the \textit{DefExtra} benchmark by implementing stricter validation, reannotation, and extension. Two authors of this work manually filtered the candidate papers and definitions to retain only author-stated definitions and exclude loose descriptions, annotator-synthesized definitions, and definitions that could not be verified in the source text. We validated definitions from retained papers and, where necessary, reannotated them from the original paper text; we kept only items on which both authors agreed. From the original $113$ papers, we kept only $60$ by this procedure. To broaden the benchmark, we further annotated $15$ additional papers, including out-of-domain papers from areas such as energy, ecology, and medicine. These additions were annotated by $11$ annotators selected under the same criteria as in TaxoMatic~\cite{Spinde2025TaxoMatic}, then checked by two authors and retained only under unanimous agreement. After validation, extension, and duplicate removal, \textit{DefExtra} contains $268$ definitions from $75$ curated papers, published between $1987$ and $2025$. All extraction and validation were conducted using Label Studio~\cite{Label_Studio}.

Each annotation contains the definition text, source metadata, a context span, and a definition-type label. The context span includes the surrounding sentences, supporting source localization and distinguishing grounded extractions from hallucinated candidates. We labeled the definitions as \textit{explicit} when they can be quoted directly from the paper, and as \textit{implicit} when a clear author-provided definition is lightly reworded while preserving its meaning, e.g., by removing irrelevant parenthetical notes. We marked out-of-domain papers in the released data (i.e., not media-bias related).

\subsection{DefSim: Definition Similarity Resource}
\label{sec:meth:defsim}

% \textit{DefSim} supports evaluation of extracted definitions by providing human judgment of definition-pair similarity. Similarity evaluation is required because LLM outputs might not match human annotations verbatim: a prediction may paraphrase the ground truth, partially overlap with it, or express a narrower concept. We construct \textit{DefSim} from $60$ definition pairs drawn from \textit{DefExtra} ``gold'' annotations and predictions of top-performing extractors described in \Cref{sec:meth:scidef}. The set contains $30$ gold--prediction pairs, $15$ gold--gold pairs, and $15$ prediction--prediction pairs, covering exact, partial, and non-matching cases.

\textit{DefSim} supports evaluation of extracted definitions by providing human judgment of definition-pair similarity. This is necessary because LLM outputs may not match human annotations verbatim: a prediction may paraphrase the ground truth, partially overlap with it, or express a narrower concept. We construct \textit{DefSim} from $60$ definition pairs drawn from \textit{DefExtra} ``gold'' annotations and top-performing extractor predictions described in \Cref{sec:meth:scidef}. The set contains $30$ gold-prediction, $15$ gold-gold, and $15$ prediction-prediction pairs, covering exact, partial, and non-matching cases.

Each pair was labeled by three authors of this paper. Annotators saw both definitions and their local contexts, and rated the conceptual similarity of the defined terms on a $1$--$5$ scale, where low scores indicate unrelated definitions, intermediate scores indicate partial overlap or scope differences, and high scores indicate semantic equivalence. We release the mean and majority human labels, along with pair metadata and our selected NLI metric score (\Cref{sec:meth:scidef}). 
% \textit{DefSim} is used as a task-specific validation resource for definition-matching metrics. 

\subsection{SciDef: Extraction and Evaluation Pipeline}
\label{sec:meth:scidef}

We release \textit{SciDef}, an extensible extraction and evaluation pipeline for scientific definitions. Given a paper, it parses the document, chunks the text, prompts an LLM to extract candidate definitions, and optionally evaluates the resulting definition set against \textit{DefExtra}. The pipeline is modular: users can replace the model backend, chunking strategy, prompt template, or evaluation metric.

We parse PDFs with GROBID~\cite{GROBID} and evaluate four paper-chunking strategies: section-level, paragraph-level, sentence-level, and a three-sentence sliding window using regex-based splitting, the latter matching the context-span format in \textit{DefExtra}. For extraction, we compare \textit{OneStep} prompting, which directly extracts definitions from a chunk, and \textit{MultiStep} prompting, which first determines whether a chunk contains a definition and extracts only if it does. We also evaluate a few-shot variant, \textit{OneStep-FS} and \textit{MultiStep-FS}, using training examples. All extractors output a definition text and an explicit/implicit type label. In addition, we implement DSPy-based versions of the one-step and multi-step extractors and optimize prompts with BootstrapFewShot, BootstrapFewShotWithRandomSearch, and MIPROv2~\cite{opsahl-ong-etal-2024-optimizing}. During DSPy optimization, models also predict the local context span as an auxiliary grounding signal (to mitigate the generation of definitions not present in the context), but context is not used during inference.

We split \textit{DefExtra} into 60\%/20\%/20\% train/development/test sets corresponding to 45/15/15 papers and $133$/$54$/$81$ definitions, respectively. The training split is used for few-shot samples and DSPy optimization, the development for prompt selection, and the test for final comparison across $16$ LLMs and prompting strategies. We evaluate both open- and proprietary-LLMs.\footnote{The full model list, prompts, provider details, and configurations are available in the released repository.}

SciDef also includes the evaluation protocol for comparing extractor outputs with \textit{DefExtra}. Since exact matching fails for paraphrased, partial, or broader definitions, we first compare candidate pairwise metrics on established semantic similarity and paraphrase benchmarks: STS3k-\{all,adv,non\}, STS-B, SICK, MSRP, and QQP~\cite{fodor-etal-2025-compositionality,marelli2014sick,doland2005paraphrase,sharma2019nlu}. We compare embedding cosine similarity, bidirectional NLI, and LLM-as-a-Judge scoring (full LLM-as-a-judge model \& prompt list available in our repository's appendix), select an NLI-based metric from these general benchmarks, and then use \textit{DefSim} as a task-specific validation resource to assess agreement with human judgments on definition pairs. \textit{DefSim} is therefore not used to tune or select the metric but rather to validate it.

For each paper, we evaluate the predicted definition set $P$ against the gold set $G$. Given the selected pairwise metric $M_{\mathrm{def}}$ (i.e., NLI-based metric), we threshold weak matches at $\tau=0.25$ and combine semantic similarity with explicit/implicit type agreement:
\begin{equation}
S(g,p)=
\begin{cases}
0, & M_{\mathrm{def}}(g,p)<\tau,\\
\dfrac{M_{\mathrm{def}}(g,p)+\mathbbm{1}[\mathrm{type}(g)=\mathrm{type}(p)]}{2}, & \text{otherwise}.
\end{cases}
\end{equation}
We then compute a bidirectional best-match score:
\begin{equation}
\label{eq:score}
\mathrm{Score}(G,P)=\frac{1}{2}\left(
\frac{1}{|G|}\sum_{g\in G}\max_{p\in P} S(g,p)
+
\frac{1}{|P|}\sum_{p\in P}\max_{g\in G} S(g,p)
\right).
\end{equation}
The first term measures coverage of human annotations, while the second penalizes over-generation. If $P$ is empty, the final score is set to zero. We also report \textit{GT coverage}, the percentage of gold definitions with at least one prediction above $\tau$.
% \footnote{Full implementation details are available in our repository.}

%% file: latex/Sections/Results.tex
\section{Resource Validation Results}
\label{sec:results}

We report validation results for all the resources. Detailed per-model/strategy results are available in the repository appendix\footnote{\url{https://github.com/Media-Bias-Group/SciDef/blob/main/APPENDIX.md}}.

\Cref{tab:dataset_examples} shows representative sample--label pairs from the two released datasets. For \textit{DefExtra}, the input contains context from a paper, and the label is the extracted term, definition, and definition type. For \textit{DefSim}, the input is a definition pair, and the label is the human similarity judgment and the similarity judged by our metric.

\begin{table*}[h]
\centering
\footnotesize
\setlength{\tabcolsep}{3pt}
\renewcommand{\arraystretch}{1.08}

\begin{minipage}[t]{0.49\textwidth}
\centering
\textbf{\textit{DefExtra}: context $\rightarrow$ extraction label}
\vspace{1mm}

\begin{tabularx}{\linewidth}{@{}>{\raggedright\arraybackslash}X>{\raggedright\arraybackslash}X@{}}
\toprule
\textbf{Sample context} & \textbf{Label} \\
\midrule
``Media bias is defined by researchers as slanted news coverage or internal bias, reflected in news articles.'' &
\textit{media bias}: ``slanted news coverage or internal bias, reflected in news articles''; explicit \\

\addlinespace[1pt]
``Publication bias arises if certain types of statistical results are more likely to be published than other results.'' &
\textit{publication bias}: ``certain types of statistical results are more likely to be published than other results.''; explicit \\

\addlinespace[1pt]
``... people often use terms that are highly offensive to certain groups but in a qualitatively different manner.'' &
\textit{offensive language}: ``Terms that are highly offensive to certain groups but are used in a qualitatively different manner''; implicit \\
\bottomrule
\end{tabularx}
\end{minipage}
\hfill
\begin{minipage}[t]{0.49\textwidth}
\centering
\textbf{\textit{DefSim}: definition pair $\rightarrow$ similarity label}
\vspace{1mm}

\begin{tabularx}{\linewidth}{@{}>{\raggedright\arraybackslash}X>{\raggedright\arraybackslash}p{1.35cm}@{}}
\toprule
\textbf{Definition pair} & \textbf{Label} \\
\midrule
\textit{publication bias}: ``certain types of statistical results are more likely to be published ...'' \newline
$\leftrightarrow$ \textit{publication bias}: ``arises if certain types of statistical results are more likely to be published ...'' &
$5/5$ \\

\addlinespace[1pt]
\textit{task planning}: ``a complex process that involves reasoning, decision-making, and coordination to ...'' \newline
$\leftrightarrow$ \textit{task planning}: ``the process of generating a sequence of actions to achieve a goal'' &
$3/5$ \\

\addlinespace[1pt]
\textit{ad hominem bias}: ``a journalist attacks another person instead of their argument ...'' \newline
$\leftrightarrow$ \textit{climatic sampling bias}: ``bias that occurs when trials are performed under unrepresentative seasonal climate conditions'' &
$1/5$ \\
\bottomrule
\end{tabularx}
\end{minipage}

\caption{Sample--label examples from the released datasets. Contexts are shortened for display. \textit{DefExtra} labels contain the extracted term, definition, and explicit/implicit type; \textit{DefSim} labels are human definition-similarity judgments on a $1$--$5$ scale.}
\label{tab:dataset_examples}
\end{table*}

\subsection{Metric Selection and DefSim Validation}
\label{sec:results:similarity}

Before running the extraction benchmark, we selected the pairwise similarity metric using established semantic similarity datasets. Across the seven benchmarks, NLI achieved the strongest performance, with a mean F1 of $0.824$, a median F1 of $0.843$, and the best score on six of seven benchmarks. Embeddings achieved a mean F1 of $0.682$ and a median F1 of $0.719$, performing best only on MSRP. LLM-as-a-Judge scoring achieved a mean F1 of $0.687$ and a median F1 of $0.713$, but did not perform best on any benchmark. We therefore use NLI as the pairwise metric for extraction scoring.

After selecting NLI, we use \textit{DefSim} to validate whether the metric transfers to scientific definition similarity. For the released definition-pair labels ($N=60$), annotators show very high agreement, with Krippendorff's $\alpha$~\cite{krippendorff2011computing} of $\alpha=0.924$, and the NLI metric aligns closely with the mean human labels with Pearson correlation of $\rho=0.937$. 
As a second validation (i.e., outside of \textit{DefSim}), we evaluate the full paper-level scoring function in \Cref{eq:score} on a separate set of 58 full-paper extraction outputs, each scored by human annotators and our automatic metric. Annotator agreement remains substantial with $\alpha=0.740$, and metric--human agreement is high with $\rho=0.918$. Together, these results support \textit{DefSim} as a task-specific validation resource for definition matching and support the NLI-based score used in the extraction benchmark.

% Similarly, we also validate the full per-paper metric in \Cref{eq:score} using $N=58$ papers with full extractions scored by our metric and show that annotator agreement with this score remains substantial with $\alpha=0.740$, and metric--human agreement is high with $\rho=0.918$.

\subsection{Extraction Benchmark with SciDef}
\label{sec:results:extraction}

\begin{table}
\centering
\footnotesize
\setlength{\tabcolsep}{2pt}
\renewcommand{\arraystretch}{1.08}
\caption{Extraction validation on \textit{DefExtra} (test set). Mean is averaged across configurations. GT, score, and prediction counts are reported for the best observed configuration in each strategy family.}
\label{tab:strategy-summary}
\begin{tabularx}{\columnwidth}{@{}l c X c c c c@{}}
\toprule
Strategy & Mean & Best Configuration & GT & Score & Pred. avg. & Pred. med. \\
\midrule
DSPy & 0.216 & Qwen3-30B-A3B-Thinking-2507-FP8, section & $69.7\%$ & \textbf{0.397} & 12.20 & 5.00 \\
MultiStep-FS & 0.219 & Magistral-Small-2509, paragraph & $66.7\%$ & 0.382 & 11.07 & 6.00 \\
MultiStep & \textbf{0.221} & Magistral-Small-2509, paragraph & $57.6\%$ & 0.341 & 9.93 & 5.00 \\
OneStep & 0.158 & gpt-5-nano, section & $77.3\%$ & 0.274 & 27.53 & 18.00 \\
OneStep-FS & 0.147 & deepseek-chat-v3-0324, section & \textbf{$86.4\%$} & 0.258 & 39.13 & 22.00 \\
\bottomrule
\end{tabularx}
\end{table}

We use \textit{SciDef} to validate \textit{DefExtra} as an extraction benchmark and to validate the NLI-based metric. \Cref{tab:strategy-summary} reports the average test score by strategy, the best observed configuration in each strategy family, its ground-truth coverage, final score, and average/median number of predicted definitions per paper. The strongest configuration is a DSPy-optimized extractor using a \textit{Qwen-family} model, reaching a test score of $0.397$. The highest ground-truth coverage is achieved by a one-step few-shot extractor, which recovers $86.4\%$ of gold definitions but obtains a lower score because it also extracts many irrelevant candidates. The mismatch between coverage, score, and prediction counts demonstrates the main empirical value of the benchmark: it distinguishes between simply finding many plausible definitions and finding the relevant definitions.

The results show that LLMs can extract definitions from scientific papers, but extraction coverage alone is insufficient. High-recall extractors often overgenerate irrelevant or non-definitional statements, while the best-scoring extractors balance coverage with specificity. Eight of the top 10 extractors are DSPy-based, suggesting prompt optimization can improve this balance. All top-10 rankings and score-vs.-count plots are available in the repository appendix.

%% file: latex/Sections/Conclusion.tex
\section{Conclusion}
\label{sec:conclusion}

We released two resources for scientific definition extraction: \textit{DefExtra}, a benchmark of human-extracted definitions from academic papers, and \textit{DefSim}, a human-labeled definition-similarity resource. To support reuse, we also released \textit{SciDef}, a reproducible LLM-based extraction and evaluation pipeline. Our validation experiments show that LLMs can recover many scientific definitions. They also show that DSPy-optimized prompting improves extraction quality and that an NLI-based metric provides a useful basis for semantic matching between predicted and human definitions.

Beyond benchmarking, the resources support practical literature-analysis workflows. Researchers can use \textit{DefExtra} to compare new extractors, \textit{DefSim} to validate definition-matching metrics, and \textit{SciDef} to collect candidate definitions for downstream tasks such as literature reviews, taxonomy construction, ontology development, and domain mapping. The practical lesson is that current LLMs are useful as high-recall definition discovery tools but should not be treated as fully automatic definition selectors: models often retrieve many plausible candidates, making relevance-aware filtering and, in high-stakes settings, human validation a target for future work.

%% file: latex/Sections/Acknowledgment.tex
% ACM wants acknowledgments in the 'acks' environment (not \section*).
\begin{acks}
This work was supported by JSPS KAKENHI JP24H00732, JST CREST Grants JPMJCR20D3 and JPMJCR2562, including the AIP challenge program, and JST K Program Grant JPMJKP24C2, Japan. It was also funded by the German Federal Ministry of Education and Research (BMBF) through the DAAD (German Academic Exchange Service), the Deutsche Forschungsgemeinschaft (DFG, German Research Foundation) project number 565115197 (BARI), and by the Humboldt Foundation.
\end{acks}

%% file: latex/Sections/GenAIDisclosure.tex
\section*{GenAI Usage Disclosure}
% Generative AI tools were used for language editing and drafting assistance. The authors reviewed and approved all final content, including claims, experiments, data, and analyses.
Generative AI systems were used in this work in four ways. First, LLMs are the subject and components of the SciDef experiments: we evaluate LLM-based extraction pipelines across models, prompting strategies, and chunking schemes. Second, in the prior TaxoMatic stage that DefExtra builds on, GPT-3.5-turbo was used for search-space expansion as described in Section~\ref{sec:meth:defextra}. Third, generative AI tools were used for language editing and drafting assistance during manuscript preparation. Fourth, generative AI tools were used for documentation drafting. All AI-assisted text, code, documentation, analyses, and claims were reviewed, corrected, and approved by the authors. No LLM-generated definition was accepted as a DefExtra gold annotation without human validation. DefSim labels were assigned by human annotators, and annotators were not instructed to defer to model or metric outputs. The released code and documentation identify AI-assisted components where applicable.